\newcommand{\ket}[1]{\mbox{$\vert #1 \rangle$}}

\newcommand{\spinup}{\ket{{\uparrow}}} 
\newcommand{\spindown}{\ket{{\downarrow}}}
\newcommand{\defin}{{\kern3pt\lower-0.3pt\hbox{:}\kern-3pt=\kern3pt}}

\documentclass[wssquare]{ws-procs975x65} 

\usepackage[pdftex]{hyperref}

\usepackage{wrapfig}
\hypersetup{
	pdfpagelabels=true,
	colorlinks=true,
	linkcolor=blue,
	anchorcolor=blue,
	menucolor=blue,
	urlcolor=blue,
	citecolor=blue,
	filecolor=blue,
	urlcolor=blue,
	pagecolor=blue,
	pdfauthor={Carsten Robens, Stefan Brakhane, Dieter Meschede, and A. Alberti},
	pdftitle={Quantum Walks With Neutral Atoms:\\Quantum Interference Effects of One and Two Particles}}

\newcommand{\figref}[2]{\hyperref[#1]{\ref{#1}(#2)}}

\begin{document}
\title{Quantum Walks With Neutral Atoms:\\Quantum Interference Effects of One and Two Particles}

\author{Carsten Robens, Stefan Brakhane, Dieter Meschede, and A. Alberti$^*$}

\address{Institut f\"ur Angewandte Physik, Universit\"at Bonn,\\
Wegelerstr.~8, D-53115 Bonn, Germany\\
$^*$E-mail: alberti@iap.uni-bonn.de\\
http://quantum-technologies.iap.uni-bonn.de/}

\makeatletter
\def\submitted#1{\par%
    \vspace*{12pt}%
    {\authorfont{\leftskip18pt\rightskip\leftskip%
    \noindent{\it Submitted to}\/:\ #1\par}}\vskip-12pt}%
\makeatother

\begin{abstract}
We report on the state of the art of quantum walk experiments with neutral atoms in state-dependent optical lattices.
%
%
We demonstrate a novel state-dependent transport technique enabling the control of two spin-selective sublattices in a fully independent fashion.
This transport technique allowed us to carry out a test of single-particle quantum interference based on the violation of the Leggett-Garg inequality and, more recently, to probe two-particle quantum interference effects with neutral atoms cooled into the motional ground state.
These experiments lay the groundwork for the study of discrete-time quantum walks of strongly interacting, indistinguishable particles to demonstrate quantum cellular automata of neutral atoms.
\end{abstract}

\submitted{Proceedings of 22nd International Conference on Laser Spectroscopy}

\bodymatter

\section{Introduction}
\begin{figure}[b]
\begin{center}
\includegraphics[width=3in]{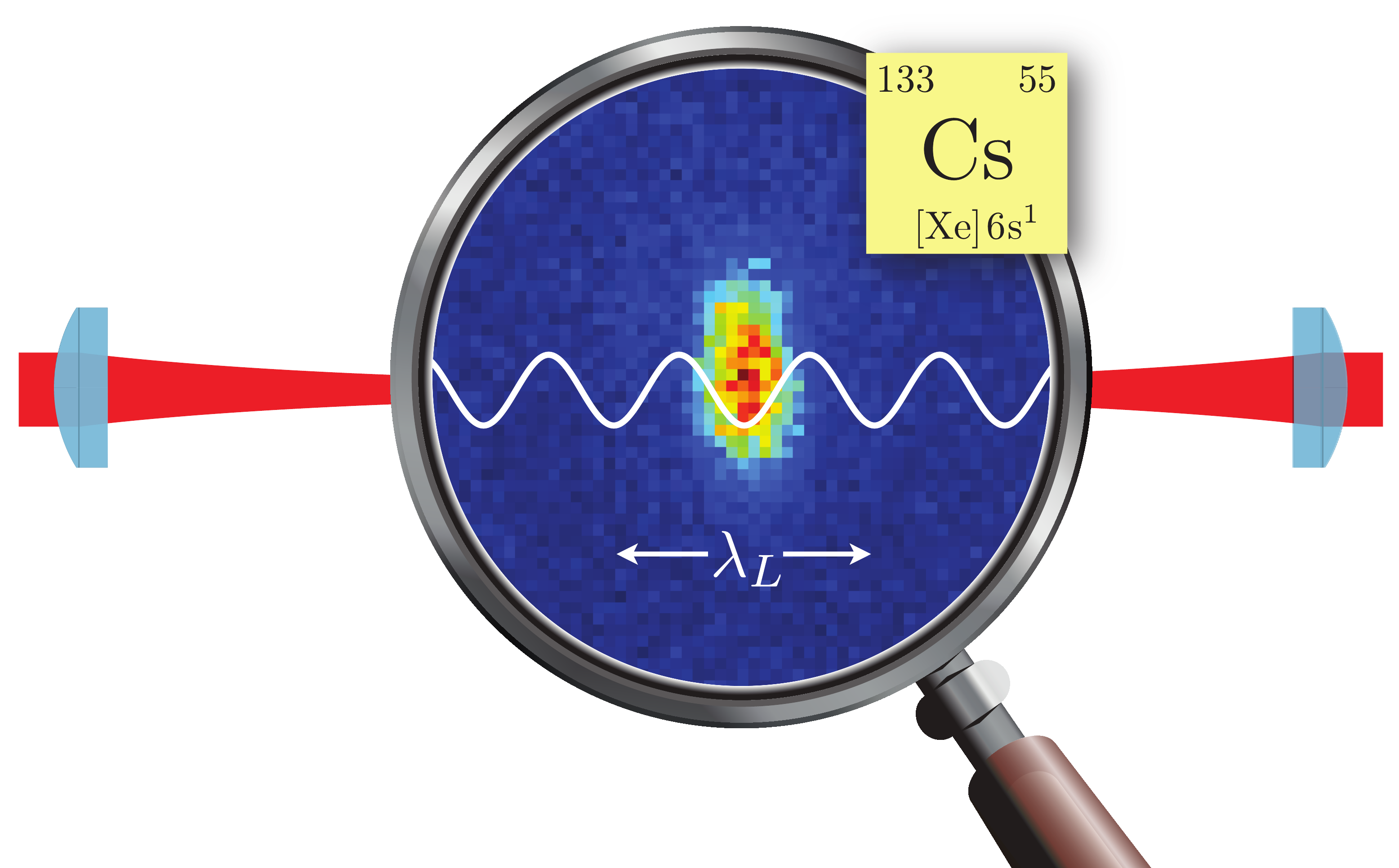}
\end{center}
\caption{Fluorescence image of a single atom in a one-dimensional optical lattice. The lattice constant is $\lambda_\text{L}/2=433\,\mathrm{nm}$. The atom image is not up to scale since the actual diffraction-limited size corresponds to four sites.}
\label{aba:fig1}
\end{figure}

The behavior of quantum particles in a periodic potential has been long investigated in physics.
These studies allowed us to understand, for instance, the motion of electrons in crystal lattices.
Since a few years, it has become possible to employ neutral atoms trapped in optical lattices to experimentally study the motion of quantum particles in periodic potentials \cite{Raizen:1997}.
The common trait of these optical lattice experiments consists in tunneling through potential barriers, which allows matter wave to coherently delocalize in space \cite{Alberti:2009}.
A different approach to achieve coherent delocalization of matter waves is provided by discrete-time quantum walks (DTQW).
Instead of continuous tunneling through barriers, DTQWs rely on the controlled motion of a quantum particle, which is rigidly shifted in discrete steps conditioned on its internal degree of freedom, constituting a pseudo spin-1/2 system.

Quantum walks hold the promise to provide a universal computational primitive \cite{Childs:2009} and are the basic building blocks of a series of quantum algorithms \cite{Shenvi:2003}. Over the past few years, experimental implementations of quantum walks have been realized with cold atoms \cite{Karski:2009} and trapped ions \cite{Blatt:2010,Schaetz:2012} with the pseudo spin-1/2 encoded in long-lived hyperfine states, as well as with photons spreading either through waveguide arrays \cite{Broome:2010} or fiber loop networks \cite{Silberhorn:2010} with the pseudo spin-1/2 encoded in the polarization states, or even different spatial modes \cite{Sansoni:2012}.

In our laboratory, we use single cesium atoms which are trapped in a very deep optical lattice potential, see Fig.~\ref{aba:fig1}.
For the experimental realization, we need to coherently control the external degree of freedom (i.e., the atom's position in the lattice) as well as the internal one constituted by the atomic spin state \cite{Karski:2009}.
Spatially resolved fluorescence detection allows us to measure the position with single site resolution \cite{Karski:2009b} and to discriminate the two spin states $\ket{{\uparrow}}$ and $\ket{{\downarrow}}$ with good fidelity via the so-called push-out method \cite{Kuhr:2005}.
More recently, we demonstrated that state-dependent optical lattices can be used to perform projective measurement of the atom's spin state even without relying on the push-out method.
This technique plays a central role in the realization of interaction-free measurement to falsify classical trajectory theories (see Sec.~\ref{sec:LeggettGarg}).

\begin{figure}[b]
\begin{center}
\includegraphics[width=3in]{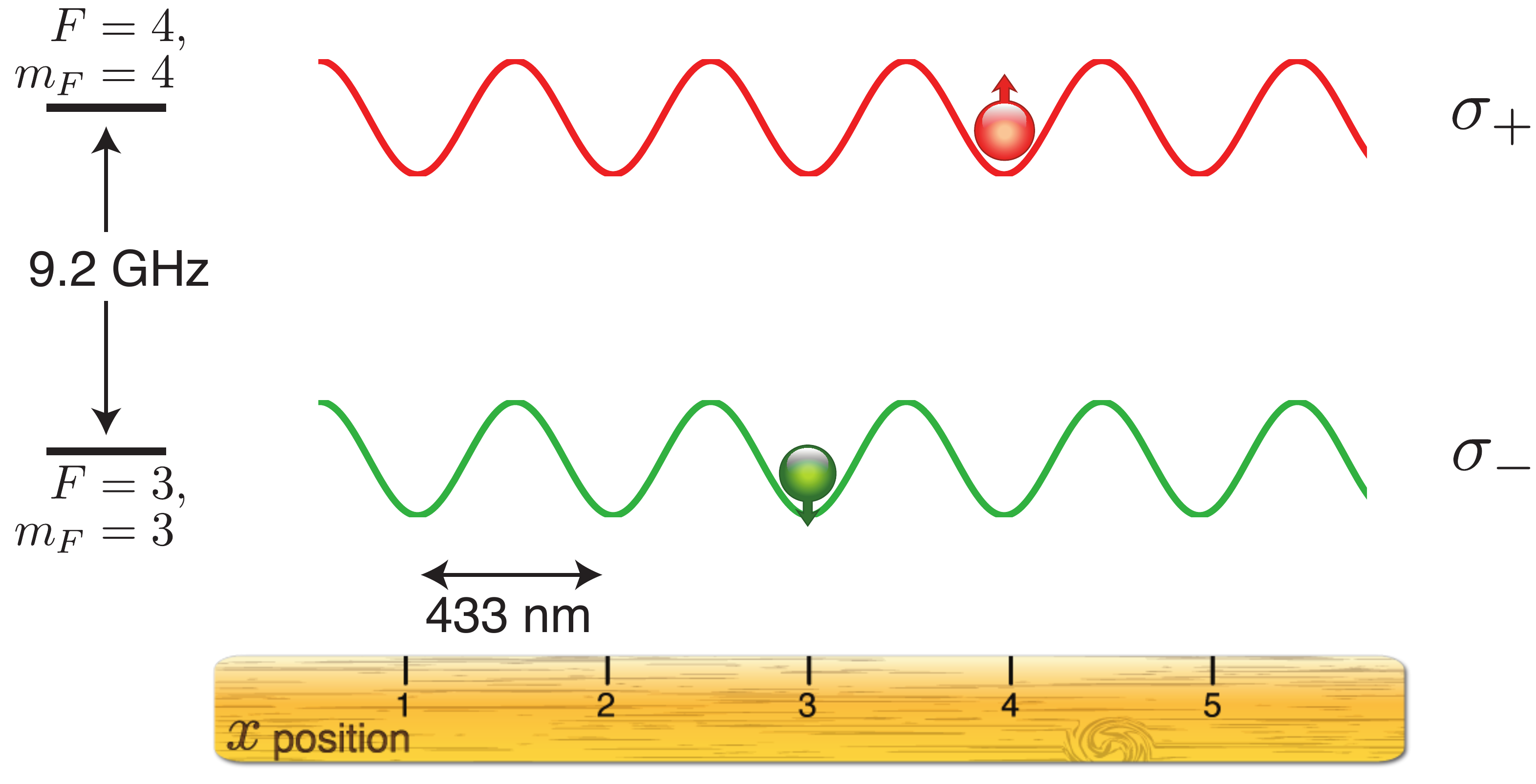}
\end{center}
\caption{State-dependent optical lattices acting selectively on either one of two long-lived hyperfine states of a cesium atom.
Upper and lower lattices originate from $\sigma_+$ and $\sigma_-$ circularly polarized standing wave light fields, respectively.}
\label{aba:fig2}
\end{figure}

\section{State-dependent optical lattice}

We use the outermost cesium hyperfine levels --- namely $\spinup \defin \ket{\mathrm{F} = 4, \mathrm{m}_{\mathrm{F}} =4}$ and $\spindown \defin \ket{\mathrm{F} = 3, \mathrm{m}_{\mathrm{F}} =3}$ --- to realize the pseudo spin-1/2 system.
The two levels can be coupled by microwave radiation at $9.2\,\mathrm{GHz}$.
Furthermore, due to the different ac-polarizability of these hyperfine levels, a magic wavelength exists at $\lambda_{\mathrm{L}}=866 \, \mathrm{nm}$ that enables spin-dependent optical potentials \cite{Zoller:1999}: Atoms in spin $\spinup$ ($\spindown$) state couple with light with $\sigma_+$ right-handed ($\sigma_-$ left-handed) circular polarization, yielding the following potentials:
\begin{equation}
	U_\uparrow(t) = U^{(0)}_\uparrow \cos\{2k_\text{L}[x-x_\uparrow(t)]\}\quad\text{and}\quad	
	U_\downarrow(t) = U^{(0)}_\downarrow \cos\{2k_\text{L}[x-x_\downarrow(t)]\}
\end{equation}
where $U^{(0)}_{\uparrow,\downarrow}$ is the lattice depth that can be individually controlled in the experiment for both spin states, as well as the individual position $x_{\uparrow,\downarrow}(t)$ of the two periodic potentials.

Since we work in a deep optical lattice, such that tunneling between lattice sites is fully negligible, the trajectory of an atom in the $\spinup$ ($\spindown$) state is determined by the motion of its $\sigma_+$ ($\sigma_-$) sublattice.
While in earlier experiments (e.g., \cite{Karski:2009,Genske:2013}) the translation of each lattice potential was realized by an electro-optical retardation plate, we have recently developed a novel method, which synthesizes the state-dependent lattices from two independent $\sigma_+$ and $\sigma_-$ optical standing waves, see Fig.~\ref{aba:fig2}.
The relative position between the two standing waves is controlled by an optoelectronic servo loop with a resolution on the order of $\lambda_L/5000$ and a bandwidth of approximately $500\,\text{kHz}$.
The bandwidth is primarily limited by the finite response time of acousto-optic modulators, which are employed to stabilize the relative position of two sublattices.
The state-dependent optical conveyor belt allows us to transport atoms arbitrarily over tens of lattice sites, as we demonstrated in one of our experiments falsifying classical trajectory theories \cite{Robens:2015} (see Sect.~\ref{sec:LeggettGarg}).

\section{Discrete-time quantum walks}
\begin{figure}[b]
\begin{center}
\includegraphics[width=4in]{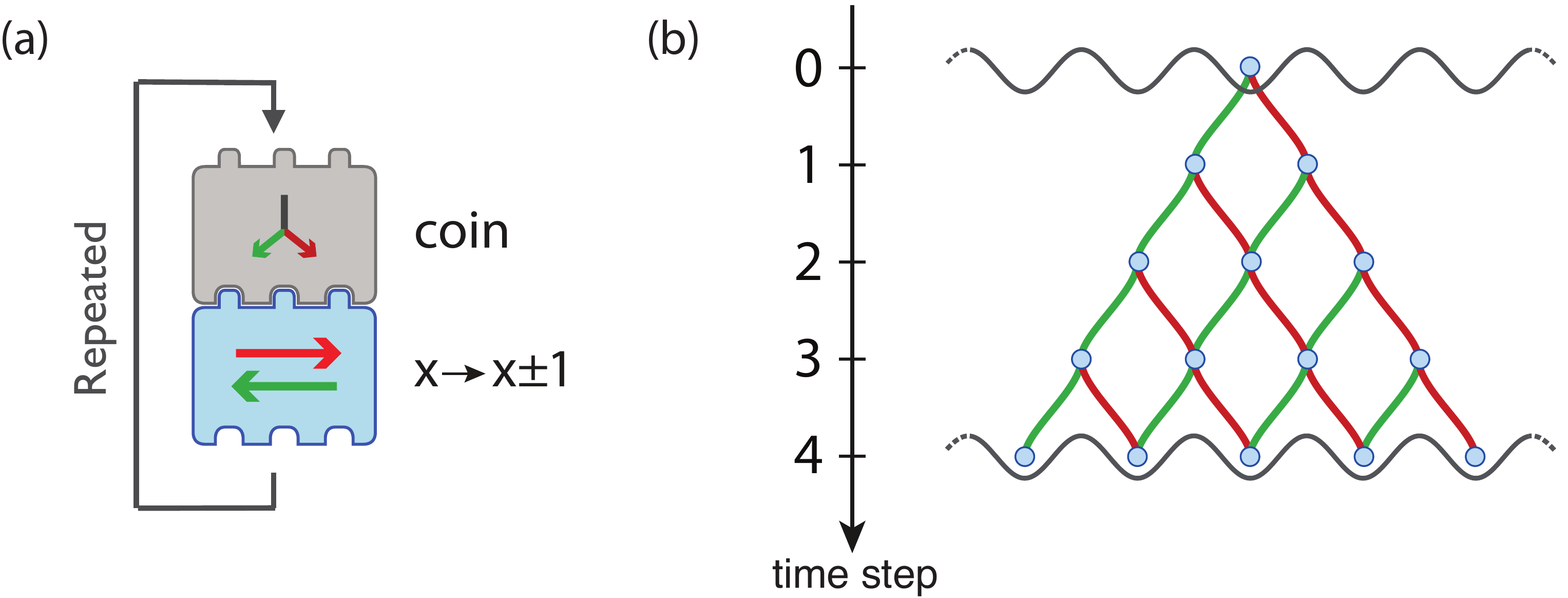}
\end{center}
\caption{Discrete-time quantum walks in position space. (a) Discrete unitary operations defining the quantum walk's step. (b) Delocalization of the quantum walker over multiple paths. The number of paths scales increases exponentially with the number of time steps.}
\label{aba:fig3}
\end{figure}
Discrete-time quantum walks are the quantum analog of random walks. In the classical world, the walker decides at discrete time steps whether to move one site leftward or rightward depending upon the result of tossing a coin --- heads or tails. As shown in Fig.~\figref{aba:fig3}{a}, a quantum walker, instead, is put at every time step in a coherent superposition of two internal states (coin operation), and it is subsequently shifted by one lattice site in a direction subject to the spin state (spin-dependent shift operation), e.g., $\spinup$ to the left and $\spindown$ to the right.

The coin operation is experimentally realized by using microwave radiation that resonantly couples the two hyperfine states.
This allows us to achieve any arbitrary unitary transformation of the pseudo spin-1/2 with the coin operation.
The most frequently used coin, however, is the Hadamard coin, which produces an equal superposition of the two spin states (coin angle equal to $\pi/2$).
The spin-dependent shift operation is realized by employing our conveyor belt transport technique, which moves the atom by one site rightward or leftward depending on the internal state.
After applying both operations the trajectory of an atom will thus be split, giving rise to a beam splitter operation of a single atom interferometer \cite{Steffen:2012}.

\begin{figure}[b]
\begin{center}
\includegraphics[width=5in]{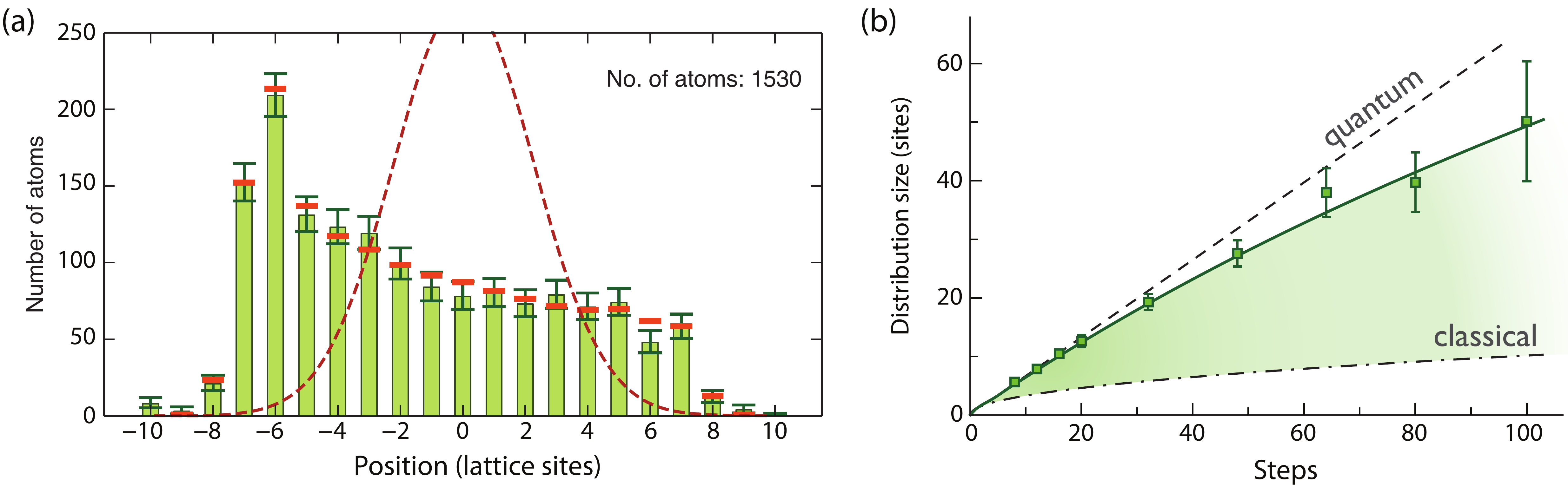}
\end{center}
\caption{Discrete-time quantum walks of single Cs atoms. (a) Probability distribution of single atoms after 20-step quantum walks. Bars with confidence intervals are the experimental data, short horizontal lines are the theoretical prediction with about $5\%$ coherence loss per step. Dashed line is the prediction for a random walk (no coherence). (b) Size of the probability distribution as a function of the number of steps. Ballistic transport (quantum) and diffusive transport (classical) are shown as asymptotic cases.}
\label{aba:fig4}
\end{figure}

After iteratively repeating the coin and shift operation, the matter wave spreads over multiple trajectories in position space, as illustrated in Fig.~\figref{aba:fig3}{b}, producing a complex multi-path interference effect.
The resulting probability distribution measured after a twenty-step quantum walk is shown in Fig.~\figref{aba:fig3}{a}, where originally the walker was prepared in site $0$ with spin $\spinup$ state.
The prominent peak on the left-hand side provides signature of multi-path interference.
Furthermore, the quantum walk spreads ballistically with the number of time steps $n$ in contrast to a classic random walk, which spreads diffusively with a Gaussian distribution of width $\sqrt{n}$.
Decoherence reduces the interference contrast, turning the quantum walk into a classical random walk.
The width of the measured probability distribution is a useful analysis tool to discriminate quantum walks from classical random walks. Fig.~\figref{aba:fig4}{b} shows the measured RMS width for an increasing number of time steps, exhibiting ballistic spreading up to a few tens of steps.
We investigated more than ten different physical decoherence mechanisms, which can be divided into two classes depending on whether they couple with the spin or positional degree of freedom~\cite{Alberti:2014}.
The short horizontal lines in the figure represent the decoherence model of quantum walks, indicating a good agreement with the experimental data.
With regard to the experimental data in Fig.~\figref{aba:fig4}{b}, the number of coherent steps is primarily limited by decoherence arising from light shifts, which however, is expected to vanish with the atoms cooled to the three-dimensional ground state of the optical lattice, see Sect.~\ref{sec:cooling}.

We exploited the possibility of delocalizing atoms over tens of lattice sites to study the physics of a charged particle, for example an electron, in a periodic potential under the effect of an external homogeneous force.
Predicted by Felix Bloch nearly 90 years ago, the quantum particle in the lattice, instead of being indefinitely accelerated, performs periodic oscillations.
We have shown experimentally that electric quantum walks do exhibit a similar behavior as Bloch oscillations, where both sublattices are accelerated at each time step for a short time interval to reproduce the action of an external electric field \cite{Genske:2013}.
However, due to the time discreteness of the electric field operation (sublattices' acceleration), an even richer range of quantum transport regimes spanning from coherent delocalization to dynamical (Anderson-like) localization has been predicted \cite{Werner:2013} and observed.

\section{Falsifying classical trajectory theories}
\label{sec:LeggettGarg}

The superposition principle is one of the pillars of quantum theory, which goes beyond the classical concept of particles moving along well defined trajectories.
To understand the motion of a quantum particle, instead, quantum theory takes into account all possible trajectories that the particle can take.
This idea lies at the heart of the quantum path integral formalism \cite{FeynmanBook}.
Up until today it is an unresolved question how to reconcile the quantum mechanical worldview, where physical objects obey the unitary Schr\"odinger equation, with the macro-realistic one, where objects are in one definite state at all times.
For instance, the macroscopic apparatus of a Stern-Gerlach experiment always measures a definite orientation of the electron's spin, although the electron is, according to quantum mechanics, in a superposition of both spin orientations.
To explain such a wave function reduction to a definite state, different ideas have been put forward, which can be coarsely divided into two groups \cite{Bassi:2013}: (a) Interpretational solutions like, among others, the decoherence approach, Bohmian mechanics, and many-worlds theory. (b) Objective collapse theories such as continuous spontaneous localization, and gravitational collapse theory.
What distinguishes theories of type (b) is the assumption of an objective reduction of superposition states involving mechanical degrees of freedom (i.e., massive particles), which deviates from a Schr\"odinger-type quantum dynamics \cite{Hornberger:2013}.

In 1985 Leggett and Garg (LG) derived an inequality relating correlation measurements performed at different times, providing an objective criterion to distinguish between quantum (a) and macro-realistic (b) theories~\cite{LeggettGarg:1985}.
The inequality is derived under two assumptions that embody the macro-realistic worldview, (A1) macro-realism (i.e., massive particles follow classical trajectories) and (A2) non-invasive measurability (i.e., the position of a macroscopic object can be measured without perturbing its subsequent evolution) \cite{Robens:2015}.
Hence, the experimental violation of LG inequality falsifies a macro-realistic description of the studied phenomenon.
Our experimental apparatus offers an ideal platform to put classical trajectories theories of type (b) to the test, as we are capable of observing the position of a massive particle, namely an atom, moving through controlled trajectories in a one-dimensional optical lattice.

\begin{figure}[t]
\begin{center}
\includegraphics[width=5in]{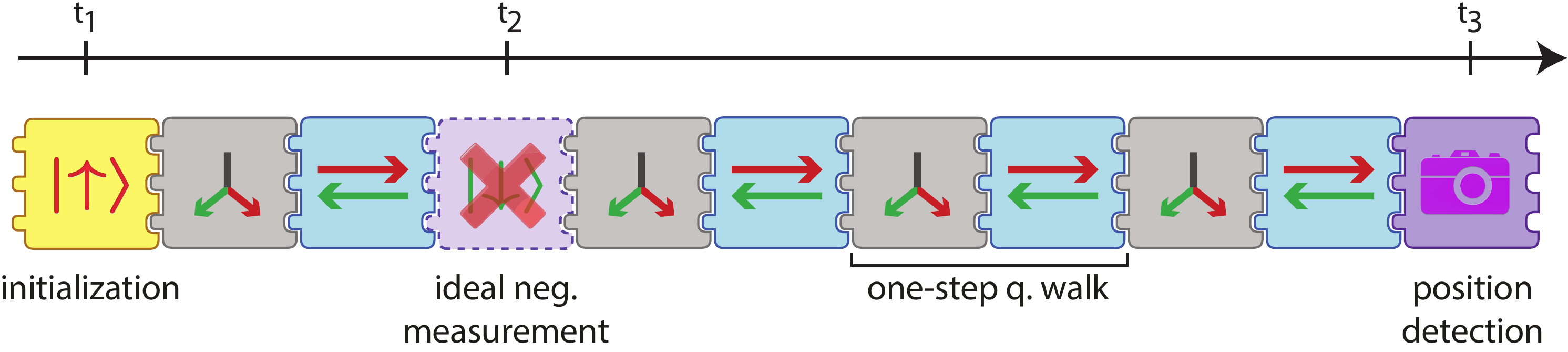}
\end{center}
\caption{Discrete-time quantum walk of four steps violating the Leggett-Garg inequality. Measurements are performed at three different times $t_i$. Only the position measurement at $t_2$ must be performed according to the ideal negative measurement protocol of Fig.~\ref{aba:fig6}. The measurement consists in spin selectively removing atoms in spin $\ket{{\downarrow}}$. Alternatively, atoms in spin $\ket{{\uparrow}}$ are selectively removed (not shown in the figure).}
\label{aba:fig6}
\end{figure}

\begin{wrapfigure}[29]{r}{2.4in}
	\vspace{-7mm}
	\begin{center}
		\includegraphics[width=2.4in]{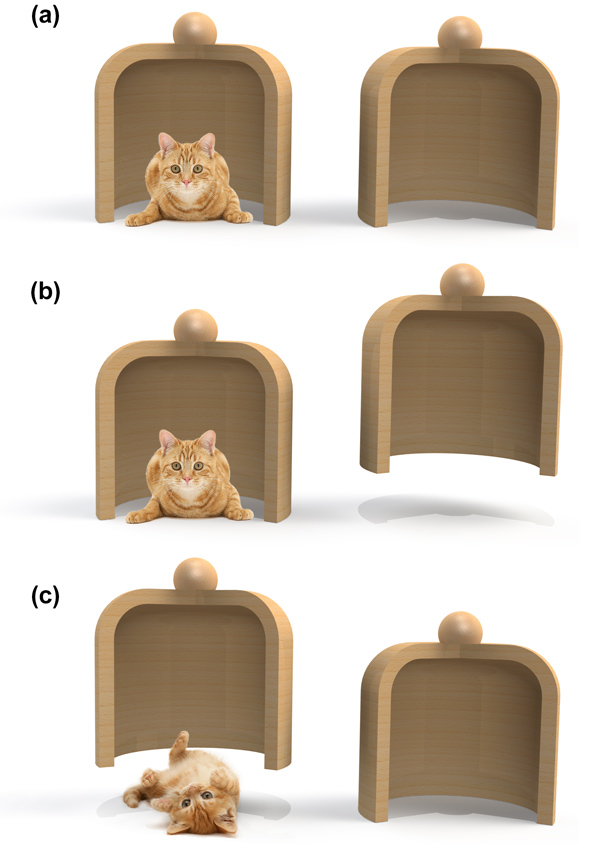}
	\end{center}
\caption{Ideal negative measurement protocol. (a) A macro-realistic cat is prepared in an unknown state either under the left or the right container. (b) Detecting the absence of the feline in the right container enables us to measure its position without any direct interaction. (c) Measurements that directly detect its presence could have disturbed the animal by direct interaction. These measurements must thus be discarded by post-selection.}
\label{aba:fig5}
\end{wrapfigure}

The LG inequality binds the linear combinations of two-time correlation measurements according to
\begin{equation}
	\label{eq:LGinequality}
	K=\large\big\langle Q(t_2)Q(t_1)\big\rangle+\big\langle Q(t_3)Q(t_2)\big\rangle-\big\langle Q(t_3)Q(t_1)\big\rangle\leq 1\,,
\end{equation}
\noindent{}where $Q(t_i)$ are measurements performed at three different times $t_i$ which are bound by $|Q(t_i)|\le  1$, but can otherwise be freely defined.
We aim to disprove macro-realistic interpretation of our DTQWs by considering a four-step quantum walk as shown in Fig.~\ref{aba:fig6}, where we choose $Q(t_i)$ to be a function of the measured particle position $x$~\cite{Robens:2015}:
\begin{equation}\label{eqn:Q1}
	Q(t_1)=+1\,,
\end{equation}
\begin{equation}\label{eqn:Q2}
Q(t_2)=\left\{\begin{array}{l}+1\quad\text{if}\quad \hat x=\;+1\\+1\quad\text{if}\quad \hat x=\;-1\end{array}\right.\,,
\end{equation}
\begin{equation}\label{eqn:Q3}
Q(t_3)=\left\{\begin{array}{l}+1\quad\text{if}\quad \hat{x}>0\\-1\quad\text{if}\quad\hat{x}\leq 0\end{array}\right.\,,
\end{equation}
\noindent{}where ${x}$ is position of the atom in units of lattice sites with respect to the initial position $x=0$ at time $t_1$.
Note that at time $t_2$ we assign the same value regardless of the measured position. This is, in fact, entirely consistent with the LG inequality and helps remark the importance that a measurement is at all performed at time $t_2$.
This measurement also represents the most challenging aspect of an experimental violation of LG inequality.
To avoid invalidating hypothesis (A2) by experimental inadvertence, Leggett and Garg themselves introduced the idea of an ideal negative measurement protocol, which is consistent with the macro-realistic hypothesis (A1).  
The idea is based on a classical measurement protocol, which consists in performing an interaction-free measurement of position as illustrated in Fig.~\ref{aba:fig5}.
While from a macro-realistic perspective, ideal negative measurements are non-invasive (hence, consistent with (A2)), it is apparent that from a quantum mechanical point of view the measurement $Q(t_2)$ causes a projection of the wave function to a definite trajectory, which in turn conditions the subsequent motion.

In the experiment, the ideal negative measurement at time $t_2$ is implemented by selectively relocating atoms in, e.g, spin $\spinup$ state so far away that they are effectively removed from the system.
Since position and spin after the first step (time $t_2$) are perfectly correlated, a measurement of spin is equivalent to one of position.
Finding the particle at time $t_3$ allows us to infer its position at time $t_2$ (in the example above: right site, $\spindown$).
We also remark that the measurement at time $t_3$ does not need to be carried out with the ideal negative measurement protocol since the atom's evolution after $t_3$ is not relevant.
Experimentally, we performed ideal negative measurements utilizing a novel spin-dependent conveyor belt technique, which allows us to shift only one sublattice at a time over multiple lattice sites, while leaving the other at rest.

\begin{figure}[h]
\begin{center}
\includegraphics[width=5in]{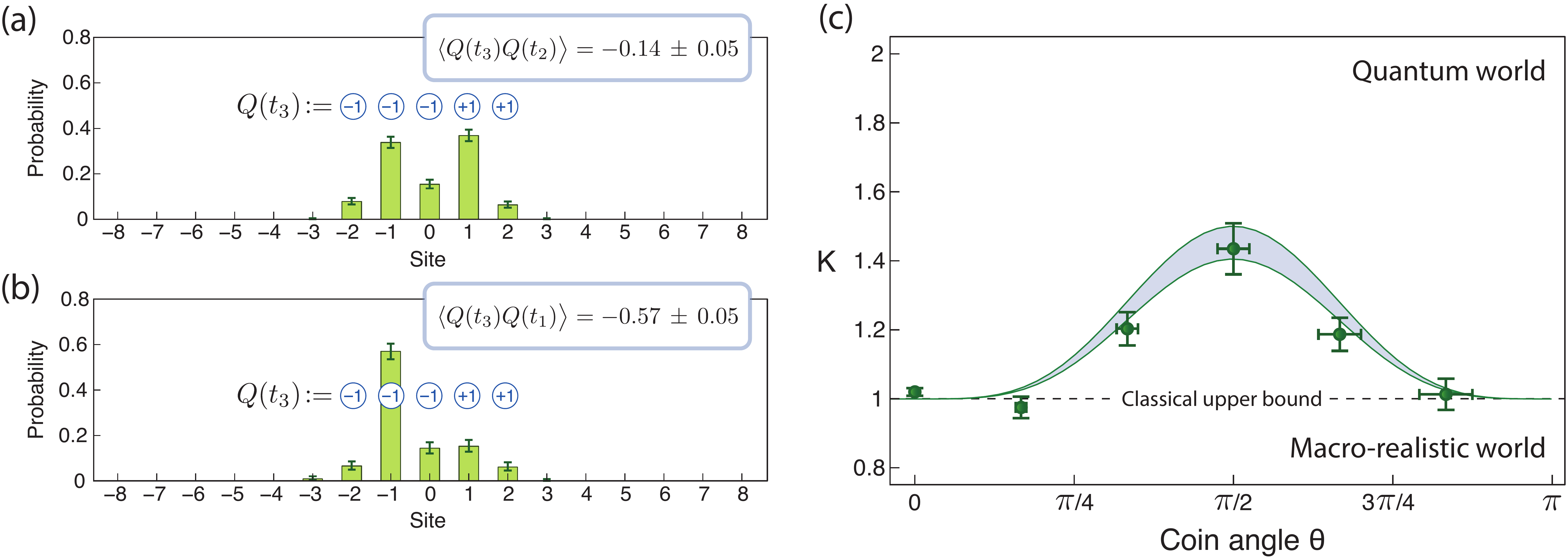}
\end{center}
\caption{Leggett-Garg test of quantum superposition principle. Measurement of the Leggett-Garg correlation functions with (a) and without (b) the ideal negative measurement at time $t_2$. (c) Maximal violation (by $6\,\sigma$) of the Leggett-Garg inequality in Eq.~(\ref{eq:LGinequality}) is measured for the Hadamard quantum walk (coin angle $\theta=\pi/2$).}
\label{aba:fig7}
\end{figure}

The measured probability distribution shown in Fig.~\figref{aba:fig7}{a,b} show distinct profiles depending on whether a measurement has been performed at time $t_2$ or not.
This difference gives rise to a violation of LG inequality when the measured probability distributions are used to compute the correlation function $K$ in Eq.~(\ref{eq:LGinequality}) 
Quantum mechanically, it is also understood that the different outcome is related to the measurement process at $t_2$, which in spite of being interaction-free, causes a projection of the wave function to a statistical mixture.
Fig.~\ref{aba:fig7} shows the value of $K$ as a function of the coin angle $\theta$, which exhibits a maximal violation of $6\sigma$ for Hadamard walk ($\theta=\pi/2$) when the coin operation prepares an equal superposition of both spin states.
Only two particular walks ($\theta=0$ and $\theta=\pi$) fulfill the LG inequality, since in these cases the coin operation creates no superposition states. 

Although our test mass --- the cesium atom --- is unquestionably microscopic, this experiment represents the most macroscopic test of quantum superposition states based on the stringent criteria provided by the LG inequality.
Many regard the LG inequality as the gold standard to discern quantum superposition states, in like manner the Bell inequality has become the widely accepted criterion to test non-locality.
Our experiment lays the groundwork for future tests with increasing ``macroscopicity'' \cite{Hornberger:2013}  that could shed light on quantum to classical transition.

\section{Three-dimensional ground-state cooling}\label{sec:cooling}
Over the years, our quantum walk experiments have demonstrated a remarkable control of single quantum particles in optical lattices.
The next frontier consists in exploring the motion of strongly correlated quantum particles \cite{Ahlbrecht:2012}.
A necessary prerequisite for these experiments is that atoms must be cooled to the lowest three-dimensional (3D) vibrational state of a single lattice site, in order to confine their motion to a very small volume and let them collide in a controllable way.

\begin{figure}[h]
\begin{center}
\includegraphics[width=3.5in]{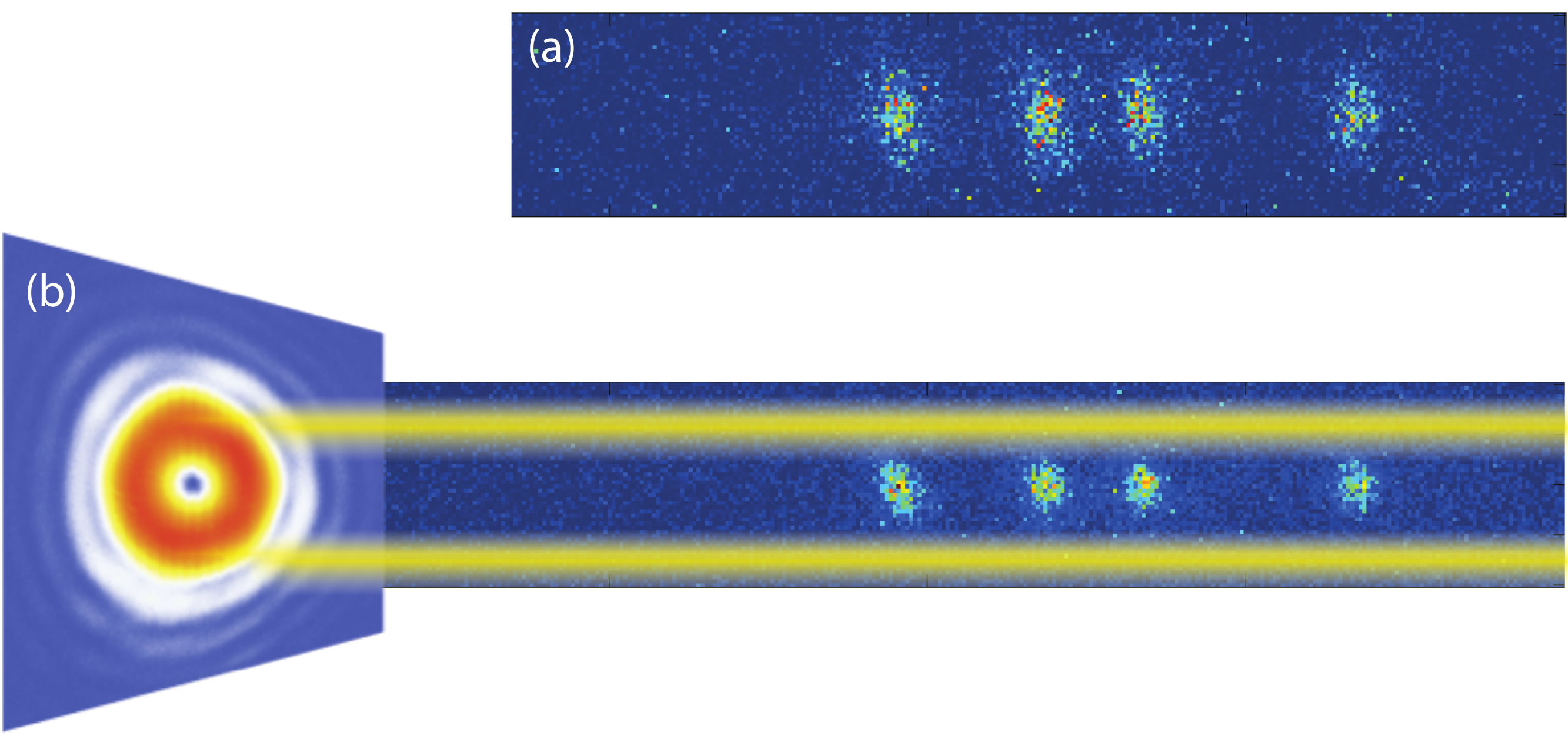}
\end{center}
\caption{Transverse compression of atoms by hollow trap. (a) Without extra transverse confinement, thermal atoms are elongated in the transverse direction ($y$-axis in the figure). (b) By adding a doughnut-shaped blue-detuned laser beam, atoms' motion is transversally squeezed. The radius  of the round shape is essentially limited by the $2\,\mu\text{m}$ optical resolution of the microscope's objective lens (NA=0.23).}
\label{aba:fig8}
\end{figure}

After molasses cooling, our cesium atoms trapped in the optical lattice occupy different 3D vibrational quantum states, with a statistical occurrence given by the Boltzmann distribution.
Yet, atoms can be subsequently prepared in the ground state by using resolved sideband cooling techniques to attain the 3D vibrational ground state \cite{Monroe:1995}.
Along the longitudinal direction of the optical lattice, we reliably employ microwave sideband cooling~\cite{Foerster:2009,Belmechri:2013}, which allows us to achieve a ground-state population of around $99\%$ starting from an initial population of around $50\%$.
Sideband cooling, however, cannot be directly employed in the transverse direction because of the weak transversal confinement with trap frequencies on the order of a few kHz (compared to the longitudinal trap frequency of $>110\,$kHz), which are on the same order or even smaller than the recoil frequency of a scattered photon ($\approx 2\,\text{kHz}$ for Cs atoms).
This corresponds to a Lamb Dicke parameter $\eta \gtrsim 1$.
To circumvent this problem we superimposed an additional blue-detuned hollow laser beam (often referred to as ``doughnut beam'') with the one-dimensional optical lattice, which increases the transversal trap frequency up to $20$\,kHz corresponding to $\eta\approx 0.3$.
Fig.~\ref{aba:fig8} shows fluorescence images of single atoms, demonstrating the transverse compression produced by the superimposed doughnut beam.

\begin{figure}[h]
\begin{center}
\includegraphics[width=\textwidth]{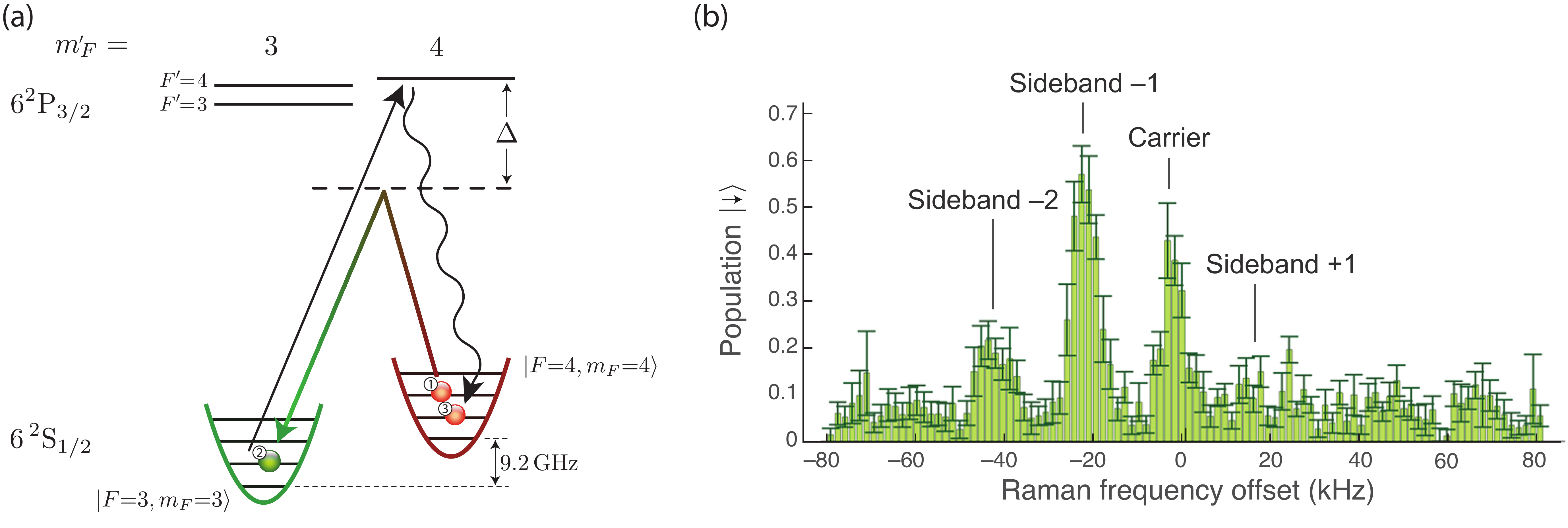}
\end{center}
\caption{Raman sideband cooling of single Cs atoms to the the 3D ground state. (a) Level scheme employed to cool Cs atoms. Circled numbers denotes the evolution of one atom during one cooling cycle. (b) Nearly-full suppression of the first blue transverse sideband indicates high occupancy of the ground state by the laser-sideband-cooled atoms. The blue longitudinal sideband at around $100\,\text{kHz}$ is also highly suppressed (not shown in the figure).}
\label{aba:fig9}
\end{figure}

We use a single pair of Raman beams (one collinear with and the other perpendicular to the optical lattice) to cool both transverse directions according to the Raman cooling scheme in Fig.~\figref{aba:fig9}{a}.
In fact, we exploit a slight ellipticity (on the percentage level) of the doughnut trap to ensure motional coupling between the two transverse directions, so that the momentum transfer provided along a single direction by the Raman transition suffices to cool the atomic motion in both transverse directions.
Fig.~\figref{aba:fig9}{b} shows an exemplary Raman sideband spectrum, where the suppression of the first blues sideband demonstrates transverse ground state cooling of atoms.
By analyzing the relative heights of the sideband peaks, we infer a transversal ground-state population of 85\% per direction, starting from an initial population of around $<1\%$. This leads to an overall 3D ground-state population of about 65\%.

\begin{figure}[h]
\begin{center}
\includegraphics[width=\textwidth]{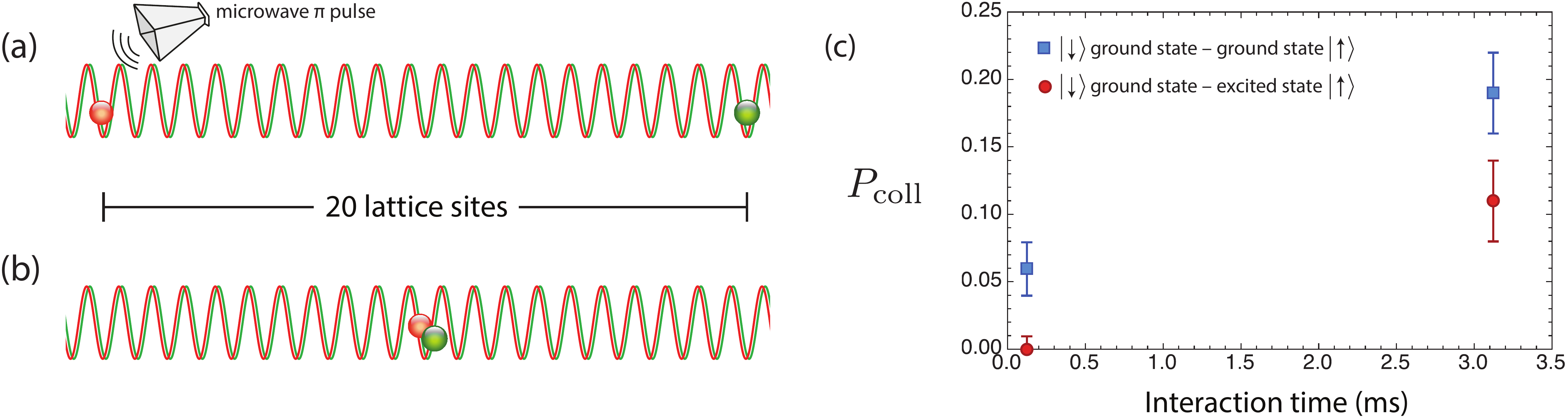}
\end{center}
\caption{Probing two-atom collisions in state-dependent optical lattices. (a) Two atoms are initially placed at a relative distance of 20 sites and cooled to the motional ground state. Using an addressing magnetic field gradient, one of the two atoms has its spin flipped by a microwave $\pi$ pulse. (b) Both atoms are transported to the same lattice site by a single adiabatic spin-dependent shift operation lasting $\approx 500\,\mu\text{s}$. (c) Based on the model in equation~(\ref{ref:collModel}), we determine the probability $P_\text{coll}$ for two atoms to be lost due to a hyperfine changing collision. By exciting selectively the atom in $\ket{\uparrow}$, we observe a reduced probability of collisional losses.}
\label{aba:fig10}
\end{figure}

\section{Probing two-atom collisions at high densities}\label{ref:atomcollisions}

The capability to control the position of individual cesium atoms with high fidelity combined with 3D ground state cooling gives us the possibility to study the motion of strongly interacting particles.
In addition, our system provides an ideal platform to measure atomic properties such as the scattering length between different spin combinations using exactly two atoms.

We carried out first experiments reporting on collisional losses due to inelastic collisions occurring at high two-atom densities.
In previous experiments~\cite{Dalibard:1998} it was shown that a dense cloud of cesium atoms undergoes hyperfine state-changing collisions on a time scale of a minute for densities $n\approx 10^{10}\,\mathrm{cm}^{-3}$.
Our experimental apparatus allows us to achieve densities six orders of magnitude higher by transporting two 3D-cooled atoms into the same lattice site according to the scheme illustrated in Fig.~\ref{aba:fig10}.

The energy released in the inelastic collision leads to losses of both atoms. By recording the occurrences at which both, one, or no atom remains in the optical lattice after a variable interaction time, we can extract the probability of inelastic collisional losses $P_\text{coll}$ using a simple model:
\begin{equation}
	\label{ref:collModel}
	\text{Probability that}\hspace{3mm}\left\{\hspace{3mm}
	\begin{array}{l}
		\text{no atom survives} = (1-P)^2\, {{P_\text{coll}}} + P^2\\[1mm]
		\text{1 atom survives} = 2P(1-P)\\[1mm]
		\text{both atoms survive} = (1-P)^2 (1-{P_\textrm{coll}})
	\end{array}
	\right.
\end{equation}
where $1-P$ is probability for a single atom to remain trapped in the optical lattice during the experimental sequence in the absence of collisions.
Independent measurements show that $1-P\approx 91\%$, which is mainly limited by technical reasons (timing of the experimental sequence) and additional losses experienced due to transverse cooling.
Experiments with tighter transverse confinement are expected to reach single-atom survival probabilities close to 99\%.

These preliminary results exhibit losses detectable already for interaction times on the ms scale, shown by the squares in Fig.~\figref{aba:fig10}{c}.
We can furthermore verify that the inelastic collision probability $P_\text{coll}$ depends on the two-atom density.
For that purpose, we excite the atom in $\spinup$ with a spin-dependent shaking of the $\sigma_+$ sublattice, which increases the volume of the atom's wave function.
The reduced probability of collisional losses in this case is shown by the circle points in the same figure.

\section{Microwave Hong-Ou-Mandel interferometer with massive particles.}
Ultracold atoms in the vibrational ground state of the lattice potential allow us to explore fascinating quantum-mechanical interference effects between two (or more) indistinguishable neutral atoms.
Quantum mechanics shows that quantum correlated states of two particles can be produced even if particles are non interacting.
The most prominent example is provided by the Hong-Ou-Mandel (HOM) experiment \cite{HOM:1987}. It demonstrated that two indistinguishable photons (with identical polarization and transverse mode) impinging simultaneously upon a beam splitter emerge in a quantum entangled state, where both photons exit from the beam splitter either through one or the other output port, but not from separate ones.
The quantum correlation results from quantum interference of two-particle trajectories, and applies in general to any indistinguishable boson particles, including massive ones.
Recently, this effect has been observed in optical tweezers \cite{Regal:2014} and atomic beams with Bose-Einstein condensates \cite{Westbrook:2015} (see Ref.~\cite{MarcCheneau2015} in this conference proceedings).

Our experimental apparatus is ideally suited to implement a direct analog of the original HOM experiment, thus demonstrating the essential building block to study correlated discrete-time quantum walks with indistinguishable particles.
Continuous-time analogues of DTQWs with correlated boson particles have similarly been observed \cite{Peruzzo:2010, Preiss:2015}.

\begin{figure}[h]
\begin{center}
\includegraphics[width=3.5in]{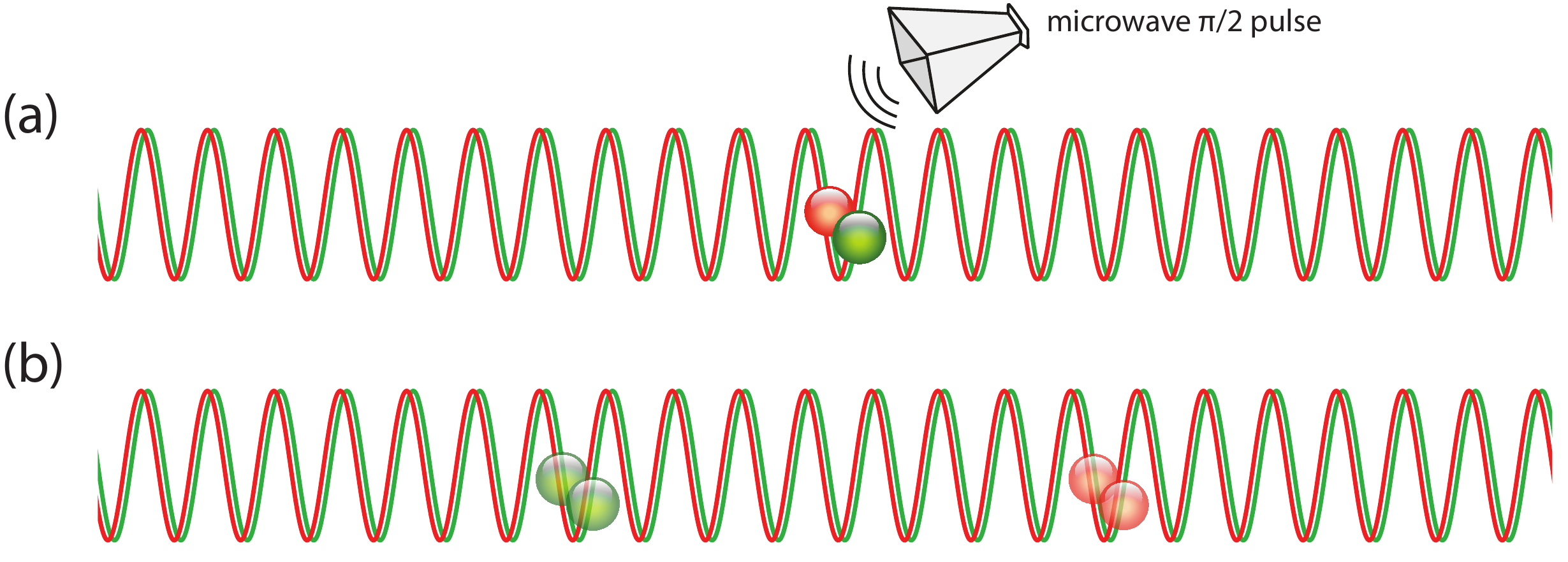}
\end{center}
\caption{Scheme of the microwave Hong-Ou-Mandel interferometer with massive particles. (a) Two atoms with opposite spin states are transported to the same lattice site in an analogous way as in Fig.~\ref{aba:fig10}. A microwave $\pi/2$ pulse mixes the two indistinguishable atoms like in the famous optical realization of the two-photon interferometer. (b) After the two spin species are separated, both atoms emerge either on the left or right hand side. For identical atoms, no event is expected with the two atoms in distinct sites.}
\label{aba:fig11}
\end{figure}

In our experimental realization of the atomic HOM effect, we initially prepare two atoms separated by 20 lattice sites, cool them into the 3D ground state, and transport them into the same lattice site using an adiabatic ramp (see Fig.~\figref{aba:fig10}{a,b}).
Instead of letting atoms interact on a ms time scale (see Sec.~\ref{ref:atomcollisions}), we directly apply a microwave $\pi/2$ pulse (see Fig.~\figref{aba:fig11}{a}) rotating the spin of both atoms onto the equator of the Bloch sphere in a time ($5\,\mu$s) much shorter than any other time scale.
By subsequently displacing our sublattices spin dependently, we expect to produce the entangled NOON state $(\ket{{\text{Right},\uparrow\uparrow}} - \ket{{\text{Left},\downarrow\downarrow}})/\sqrt{2}$, as shown in Fig.~\figref{aba:fig11}{b}.
Analogously to the original detection method \cite{HOM:1987}, we provide experimental signature of the two-particle interference by recording the suppression of events where two atoms are detected at different lattice sites, corresponding to an anti-bunched state.
Ideally these events should not occur, yet they are detected in the experiment due to imperfect ground state cooling, which impairs the indistinguishability.
In addition, our detection method does not allow a direct, unambiguous identification of all physical events due to technical reasons (finite survival probability and imperfect efficiency of parity projection \cite{Preiss:2015}). Therefore, we need to resort to a Monte Carlo analysis of our results relying on experimental parameters measured independently.
Our analysis yields a signature of the HOM interference with a statical significance of about $3\,\sigma$.
In addition, the Monte Carlo analysis confirms that the observed suppression of events with anti-bunched atoms is compatible with a 3D ground state population of around $60\%$.
A detailed investigation of systematic effects affecting our Monte Carlo analysis will be the subject of future work.

\section{Conclusions}
We reviewed the state of the art of quantum walk experiments in state-dependent optical lattices.
The ability to transport atoms in a spin dependent fashion through two fully independent sublattice potentials opens new ways to study the physics of two or few quantum particles.
One long-term goal consists in realizing quantum cellular automata of interacting, indistinguishable particles.

The challenge for the future is to extend this technology from one-dimensional to two-dimensional lattices.
This will allow us to study, in particular, new topological phases, where atomic matter waves propagating in a unidirectional fashion along the boundary of a topological island are expected to manifest \cite{Asboth:2015}.

\section*{Acknowledgements}
The authors gratefully thank Wolfgang Alt and Jean-Michel Raimond for insightful discussions, and Gautam Ramola for helpful contributions.
The authors also acknowledge financial support from NRW-Nachwuchsforschergruppe ``Quantenkontrolle auf der Nanoskala'', ERC grant DQSIM, EU project SIQS. In addition, CR and SB from BCGS program and CR from Studienstiftung des deutschen Volkes.

\bibliographystyle{ws-procs975x65}

\end{document}